# When AI Data Center Dynamics Shift: Spectral Changepoint Detection in Ambient PMU Data

Chetan Mishra, Jaime Delaree Jr. and Kevin Jones
ET Strategic Initiatives, Dominion Energy,
Glen Allen, VA, USA
chetan.mishra@dominionenergy.com

Luigi Vanfretti
Department of ECSE, Rensselaer Polytechnic Institute,
Troy, New York, USA
vanfrl@rpi.edu

*Abstract*— **Power system oscillatory dynamics often undergo significant variations, which are directly reflected in ambient data. These variations challenge traditional monitoring tools by continually shifting the baseline of normal behavior. AI training loads, which abruptly change their consumption characteristics every few days, are a prominent example. This paper proposes a data-driven method that segments ambient data into homogeneous intervals by their spectral characteristics and validates it on synthetic and real PMU data.**

## I. Introduction

The critical challenge with large-scale artificial intelligence (AI) training loads is their unique, highly dynamic footprint on the power grid, which sets them apart from both traditional electromechanical loads and standard cloud computing infrastructure. A substantial share of their demand is typically served through high-capacity power electronic converters and UPS systems, although the exact topology, including any on-site rotating generation, depends on the facility design. Under weak grid conditions, controls in the data center's power apparatuses can introduce fast oscillations [1]. Furthermore, when local grid operational thresholds are crossed, these systems can execute voluntary load transfers to backup generation, effectively appearing as sudden multi-gigawatt load losses that existing ride-through practices do not yet address [2]. While conventional industrial and residential loads follow predictable, slow-moving diurnal cycles, AI training workloads are characterized by massive, rapid power fluctuations driven by alternating computational and communication phases [3]. This fast, periodic cycling of large GPU clusters can induce sudden power swings of hundreds of megawatts within seconds, which can threaten transient stability and challenge real-time balancing [4] [5].

Lacking adequate models of and insight into this load behavior, utilities increasingly rely on synchrophasor measurements to characterize it and detect unforeseen events. Existing AI training load modeling efforts [5] largely treat the load as a stationary process. However, because AI clusters routinely change their model scheduling, parallel processing algorithms, and training configurations, their underlying electrical and oscillatory characteristics can alter completely every few days. This was shown in our prior work [6], which analyzed more than a year of synchrophasor data at an AI training load to characterize real-world data center load behavior. Fig. 1 plots a two-month subset, illustrating how specific operating modes recur over days or weeks.. Segmenting data into homogeneous intervals enables analyzing and modeling these recurring regimes, such as periodic computational phases or reused training configurations.

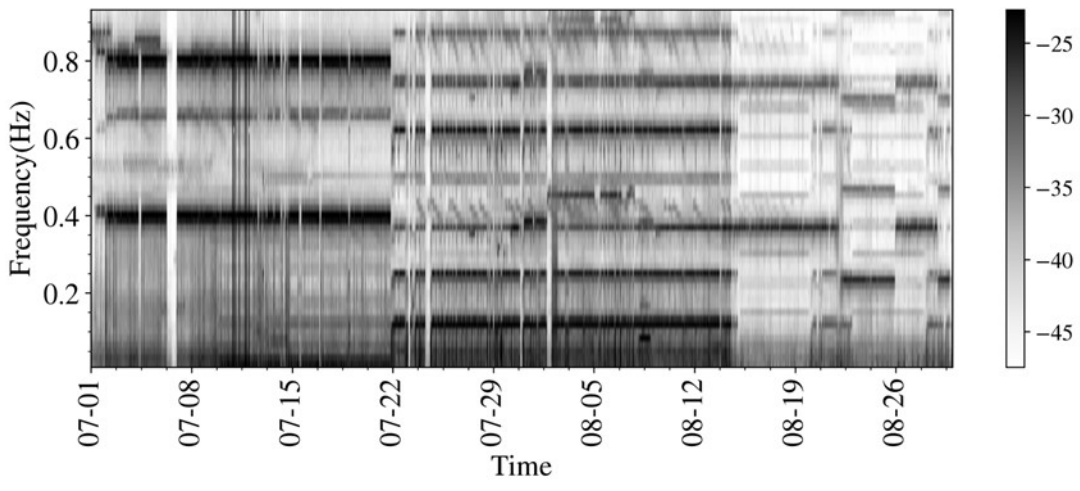

Fig. 1. Low-frequency (< 1 Hz) current spectrogram of an AI training load for two months, showing its changing spectrum.

The approach is not limited to AI training loads; it also applies to wide-area monitoring more broadly. Wide-area monitoring tools were traditionally focused on analyzing inter-area oscillations [7], which involve a large number of assets, which makes their characteristics robust to minor changes in operating conditions. Repurposing these tools for local dynamics, however, is challenging because the baseline of normal operation shifts. For example, an oscillatory mode associated with solar PV plant control will only be visible during periods of high irradiance (daytime) [8]. Standard monitoring frameworks rely on static statistical thresholds, leaving them vulnerable to high false-alarm rates when baseline grid behavior shifts [9]. Automatically identifying these changepoints isolates recurring operating patterns and enables adaptive monitoring.

Prior synchrophasor change-detection methods largely focus on detecting transients [10], [11], exploiting low ranked nature of trend/ operating point. The work in [12] is closest in spirit to our work, where a changepoint detection framework is adopted to detect onset of a single narrow band oscillation, modeling it as a univariate mean shift problem with stationary ambient dynamics. However, emerging regime shifts in AI training loads and converter controls manifest as wideband transitions across both the continuous stochastic ambient spectrum and multi-harmonic duty cycles, where the active frequencies are unknown *a priori* and the ambient baseline itself shifts. Directly applying changepoint detection across a full spectral estimate, however, is hindered by the curse of dimensionality, non-Gaussian periodogram noise, and covariance identifiability issues. The primary contributions of this paper are:

1. We formulate a wideband spectral segmentation framework that captures simultaneous regime shifts in the ambient spectrum and load duty cycles.
2. Exploiting the inherent low-rank nature of real-world spectral changes under a mixed-spectrum model, we establish that applying PCA to log-spectra compresses both ambient noise and deterministic oscillations into uncorrelated, low dimensional Gaussian features, enabling a tractable diagonal-covariance GLR-BIC changepoint detector.
3. Extensive Monte Carlo simulations on synthetic data alongside Dominion Energy PMU measurements from an AI training load and a data center UPS oscillation event confirm the framework's precision and practical utility.

The paper is organized as follows. Section II formulates the problem and spectrally characterizes ambient data. Section III presents the proposed approach, and Section IV presents results on synthetic data and on real-world Dominion Energy synchrophasor data.

## II. Ambient Data Characterization and Problem Formulation

To capture abrupt changes as seen in Fig. 1, we adopt a piecewise stationary signal model for ambient data. Consider a synchrophasor data window $\{x[k], k = 0,1, \dots K-1\}$ sampled at $f_s$ Hz ($\Delta t = \frac{1}{f_s}$ interval). If it consists of $M$ homogeneous, non-overlapping intervals, partitioned by unknown changepoint indices $0 = k_0 < k_1 \dots . < k_M = K$, the measurements within each segment can be modeled as,

$$x[k] = \epsilon_m[k], \qquad k \in (k_{m-1}, k_m] \ \forall m \tag{1}$$

where $\epsilon_m[k]$ represents a combination of periodic components (load duty cycle) and stochastic ambient dynamics (modes, measurement noise, etc.) in the $m^{th}$ segment. This can be modeled as a single realization of a stationary process with a mixed spectrum [13],

$$\epsilon_m[k] = \int_{-\frac{f_s}{2}}^{\frac{f_s}{2}} e^{j2\pi f k \Delta t} dZ_m(f) \tag{2}$$

where $E[dZ_m(f)]$ captures the deterministic oscillatory component at $f$ Hz and $COV(dZ_m(f), dZ_m^*(f_1)) = S_m(f)\delta(f - f_1)df$, with $S_m(f)$ being the power spectral density (PSD) of ambient noise in the $m^{th}$ interval.

Spectral characterization is used because frequency-domain analysis techniques [13] have proven effective for analyzing real-world power system dynamics [1]. Oscillatory dynamics (modes) usually occupy distinct frequencies, so the analysis can isolate and target the dynamics of interest in the frequency domain. Furthermore, being non-parametric, these techniques avoid restrictive time-series modeling assumptions.

Because we aim to detect variations in both ambient dynamics and periodic components, a time-localized estimate of average signal power serves as an effective feature space. A spectrogram $\hat{S}[i,j]$ (e.g. Fig. 1) acts as a baseline estimator for average process power around the $i^{th}$ DFT frequency bin within the $j^{th}$ time interval, defined by time index $k \in [jH, jH + N)$,

$$\hat{S}[i,j] = \left| \frac{\sum_{n=0}^{N-1} x[jH + n] e^{-j2\pi n \frac{i}{N}}}{\sqrt{N}} \right|^2 , j = 0,1 \dots \tag{3}$$

Conceptually, the above operation slides an $N$ length window across the data $x[k]$ with a hop size of $H$ points to extract time-localized segments, followed by mapping them into the frequency domain through a Discrete Fourier Transform (DFT). Because $\hat{S}$ is computed over finite intervals, a true changepoint at time index $k_m$ is mapped to the discrete index $j$ whose corresponding interval center $(j_m H + \frac{N}{2})$ is closest to $k_m$.

Note that (3) is a naïve estimate of average process power, with more sophisticated ones usually involving some form of smoothing over time-frequency [13]. An added advantage of this representation is that it allows us to focus on detecting changes in dynamics of interest (by choosing appropriate rows of $\hat{S}$), e.g., below 1 Hz to capture changing electromechanical modes. We therefore assume $\hat{S} \in R^{I \times J} > 0$, where $(I, J)$ are in accordance with the technique used as well as the band of interest, for the same input signal of length $K$. Note that this specific representation omits phase information. This choice is justified because ambient dynamics and noise are largely characterized by process power rather than phase. Furthermore, regarding the deterministic oscillatory components, phase changes have little bearing on their impact on the power system.

## III. Proposed Methodology

### A. Embedding Spectra

Directly using the spectral estimate $\hat{S}[:,j]$ to represent time segment $j$ is challenging because the feature dimension $I$ scales with the window length $N$ (from (3)). For instance, a 2-minute PMU window sampled at $f_s = 30$ Hz yields $I = 1800$ frequencies up to the Nyquist limit. In practice, grid events typically alter only a narrow subset of underlying dynamics. For example, in Fig. 1, the change on 7/22 is marked by the 0.4 Hz and 0.8 Hz components vanishing. Consequently, unperturbed frequency bins contribute only to degrading statistical identifiability of changepoints and inflating computational cost.

In speech processing, spectral dimensionality is reduced via Mel-frequency cepstral coefficients (MFCCs) [14]. Because speech consists of periodic excitations shaped by a smooth vocal-tract transfer function, taking the logarithm makes these components additive, and projecting onto a Fourier/cosine basis separates the smooth envelope from harmonic spikes. Real-world grid log-spectra exhibit an analogous composite structure [8]: a smooth ambient baseline driven by stochastic load variations [15] superimposed with sparse oscillatory peaks. However, whereas human speech occupies fixed physiological frequency bands that permit predefined Mel-filter binning, inverter-dominated grids lack a static spectral template, as wideband dynamics can emerge at arbitrary frequencies [8].

Fortunately, segmentation requires capturing only the *subspace of spectral variation* rather than reconstructing the full spectrum. A data-driven reduction is feasible because temporal variations in the true spectrum $S[i,j]$ are typically strongly correlated across frequencies $i$. This is because individual measurement channels primarily observe localized dynamics from nearby assets, each of which shapes a distinct frequency band (specific rows of $S$). Because these dynamics respond similarly to grid-side perturbations, strong statistical dependencies emerge across rows of $S$. Furthermore, a change point induced by, e.g., an asset disconnection removes all its associated control modes, inducing synchronized power shifts across those bands.

An added advantage of dimensionality reduction is that it enhances robustness against minor deviations from the ideal piecewise-stationary model assumed in Section II. For example, system modes undergo normal diurnal variations due to continuous changes in operating conditions. Projecting onto the dominant low-dimensional subspace suppresses these continuous ambient drifts, so that the detector responds mainly to abrupt structural transitions.

Motivated by the MFCC framework, we begin by taking the element-wise logarithm of $\hat{S}[i,j]$, a transformation well-suited for linear subspace reduction. Its statistical properties follow directly from the asymptotic (in window length $N$) characterization of $\hat{S}[i,j]$ in (3) for a Gaussian process $x[k]$ [13],

$$\hat{S}[i,j] \sim \frac{S[i,j]}{2}\chi_2^2\left(\lambda = \frac{A^2[i,j]N}{2S[i,j]}\right) \qquad (4)$$
$$\hat{S}[0,j] \perp \hat{S}[1,j] \dots \perp \hat{S}[I-1,j]\ \forall j$$

where $\chi_2^2(\lambda)$ denotes a non-central chi-square distribution with 2 degrees of freedom and non-centrality parameter $\lambda$, while $A[i,j]$ and $S[i,j]$ are the oscillation amplitude and ambient noise PSD at the $i^{\text{th}}$ frequency in the $j^{\text{th}}$ time block of length $N$, respectively. Under a purely ambient regime with no deterministic oscillations ($A[i,j]=0$), applying a logarithmic transform to (4) separates the unknown deterministic true spectrum $\log(S[i,j])$ from an additive noise term $\log\left(\chi_2^2(0)\right)$ that is i.i.d. across $(i,j)$. This variance-stabilizing transformation equalizes the estimation error variance across all time-frequency bins, preventing frequency indices $i$ with large baseline power $S[i,:]$ (and consequently high estimation variance) from dominating linear dimensionality reduction.

Finally, Principal Component Analysis (PCA) [11] is applied to $\log(\hat{S})$ to extract $r \ll I$ principal features $X$.

$$Row\ Centered\left(log(\hat{S}) \in R^{I\times J}\right) = log(\hat{S}) \times \left(I_J - \frac{1}{J}11^T\right) \qquad (5)$$
$$= U \times \Sigma \times V^H$$
$$X = V^H[1{:}r,:] \in R^{r\times J}$$

where $I_J$ denotes the $J \times J$ identity matrix.

## B. Changepoint Detection Cost Function

From (4) and (5), the projected features $X[p,j]\forall p$ are mutually uncorrelated, as they are formed by orthogonal projections (weighted by the columns of $U$) of independent random variables $\log\left(\hat{S}[i,j]\right)\forall i$ as seen in (5). Furthermore, while a Gaussian approximation for individual $\log\left(\hat{S}[i,j]\right)$ bins is already widely exploited in spectral analysis [13], the stochastic component of $X[p,j]$ linearly aggregates $I$ i.i.d. $\log\chi_2^2\ (0)$ variables (from (4)); consequently, the Central Limit Theorem (CLT) drives $X[p,j]\forall p$ much closer to a Gaussian distribution [16]. Now, for frequency bins containing deterministic oscillations ($A[i,j] \neq 0$), a first-order Taylor expansion shows that $\mathrm{Var}(\log(\chi_2^2(\lambda))) \approx 4/\lambda$, where $\lambda = \frac{A^2[i,j]N}{2S[i,j]}$ (from (4)). Because this variance decays as $\mathcal{O}(1/N)$, for large $N$, these oscillatory bins contribute negligibly and do not violate the normality assumption. Accordingly, at each time block $j$, we model $X[:,j] \sim \mathcal{N}\left(\mu_j, \Sigma_j\right)$ with diagonal covariance $\Sigma_j = \mathrm{diag}\left(\sigma_{1,j}^2, \dots, \sigma_{r,j}^2\right)$. In this reduced $r$-dimensional subspace, testing for a single changepoint at $j_1$ reduces to a multivariate Gaussian model selection problem:

$$\mathcal{M}_0(Simple): X[:,0{:}J-1] \sim N(\mu, \Sigma) \qquad (6)$$
$$vs.$$
$$\mathcal{M}_1(Model): \begin{cases} X[:,j] \sim N(\mu_1, \Sigma_1)\forall j \in [0, j_1-1] \\ X[:,j] \sim N(\mu_2, \Sigma_2)\forall j \in [j_1, J-1] \end{cases}$$

Treating the covariance $\Sigma_j$ as unknown allows the framework to accommodate general spectral estimators (e.g., Welch or multitaper). More importantly, at frequencies with deterministic oscillations, the log-variance depends explicitly on both the ambient PSD $S[i,j]$ and the oscillation amplitude $A[i,j]$ (via $\lambda$), unlike purely ambient frequencies where the log-variance is constant.

The Generalized Likelihood Ratio (GLR) framework [16] is widely used for this problem, selecting the more complex model $\mathcal{M}_1$ only if it yields a statistically significant increase in the maximized likelihood function. Note that because the unconstrained model $\mathcal{M}_1$ is strictly more flexible (having twice the number of free parameters, $4r$ vs. $2r$), splitting a segment never decreases the maximized likelihood, so an unpenalized criterion always favors spurious splits. To prevent this, we adopt the Bayesian Information Criterion (BIC) to penalize model complexity in proportion to the degrees of freedom added by each split [16]. This penalized likelihood formulation naturally extends to the multiple-changepoint detection problem [17], where the total objective (to be minimized) decomposes additively over candidate stationary segments $[j_1, j_2]$ via the penalized segment cost,

$$c(j_1, j_2) = \underbrace{(j_2 - j_1 + 1)\, log\, det\left(\hat{\Sigma}(j_1, j_2)\right)}_{2 \times neg\, log\, likelihood} + \underbrace{2r}_{\substack{number\ of\ params \\ to\ estimate}} \; log(J) \qquad (7)$$

$$Total\ Cost(0 = j_0 < j_1 \dots j_M = J) = \sum_m c(j_m, j_{m+1})$$

where $\hat{\Sigma}(j_1, j_2)$ is the maximum likelihood estimate of the diagonal covariance matrix using the samples $\{X[:, j], j_1 \leq j \leq j_2\}$ [16]. The total cost in (7) can be minimized with the algorithms in [17], to get the locations of the changepoints.

## IV. Results

The signal sampling rate is set to $f_s = 30$ Hz, typical of PMUs, with a spectrogram window length of $N = 1800$ samples (60 s) with a standard hop length $H = \frac{N}{2}$ (30s), yielding $\hat{S}[:, j] \in R^{I=900} \forall j$. Note that this is an offline segmentation framework, as PCA is computed across the full $J$-window record. The subspace dimension $r$ is selected via the singular-value elbow of $\Sigma$ in (5). We minimize (7) using the PELT algorithm [17], with BIC inspired penalty $\beta = 2r\log(J)$. The minimum segment length is set to 1 block in synthetic tests and 20 blocks (10 minutes) in field PMU tests to filter out transient changes.

### A. Synthetic Data

A two-hour window of piecewise-stationary synthetic ambient data (in accordance with Section II) is generated as the output of a linear dynamical system driven by zero-mean Gaussian white noise plus additive periodic components and measurement noise. A single changepoint is introduced around the 2984 second mark, at which the system's frequency response abruptly changes. Fig. 2 illustrates the spectrogram for a single realization of this system, where the red vertical line denotes the index $j_1 = 98$ (centered at 2970 seconds), which the true changepoint time instance is mapped to. We generate 1,000 independent realizations of this system. Applying PCA to $\log(\hat{S})$ reveals two dominant principal components ($r = 2$) in Fig. 3 (a), with the resulting, faithfully reconstructed log-spectrum (using (5)) shown in Fig. 3(b).

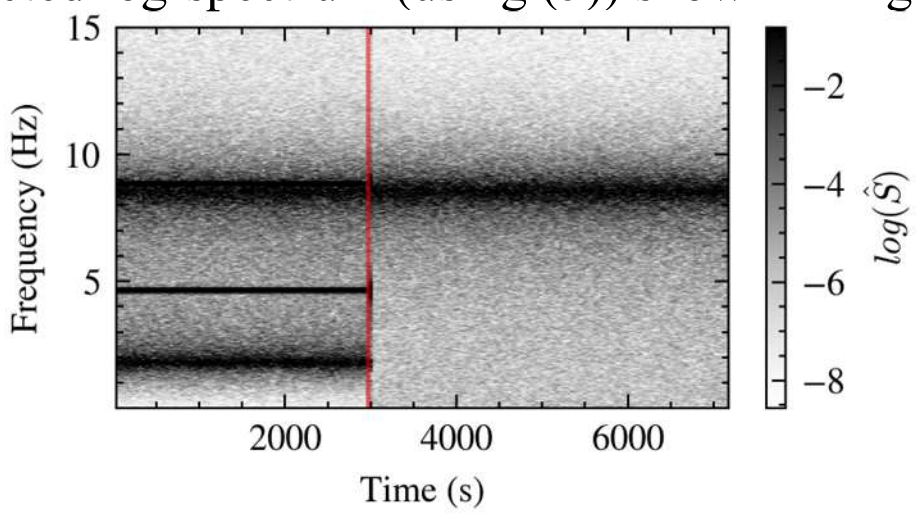

Fig. 2. Spectrogram $\hat{S}$ of synthetic ambient power system data featuring a single abrupt changepoint at $t = 2984\ s$ (red vertical line).

Next, we test the normality assumption at an arbitrary time frame ($j = 0$) by computing the coefficient of determination ($R^2$) from a linear fit against standard normal $\mathcal{N}(0,1)$ quantiles (Fig. 4). As shown in Fig. 4, across all $I = 900$ frequency bins (boxplots), the median $R^2$ increases from 0.82 for raw $\hat{S}[:,0]$ to 0.94 for $\log(\hat{S}[:,0])$, while the PCA features $X[1:r, 0]$ achieve $R^2 \geq 0.99$, validating the asymptotic modeling assumptions in Section III.A.

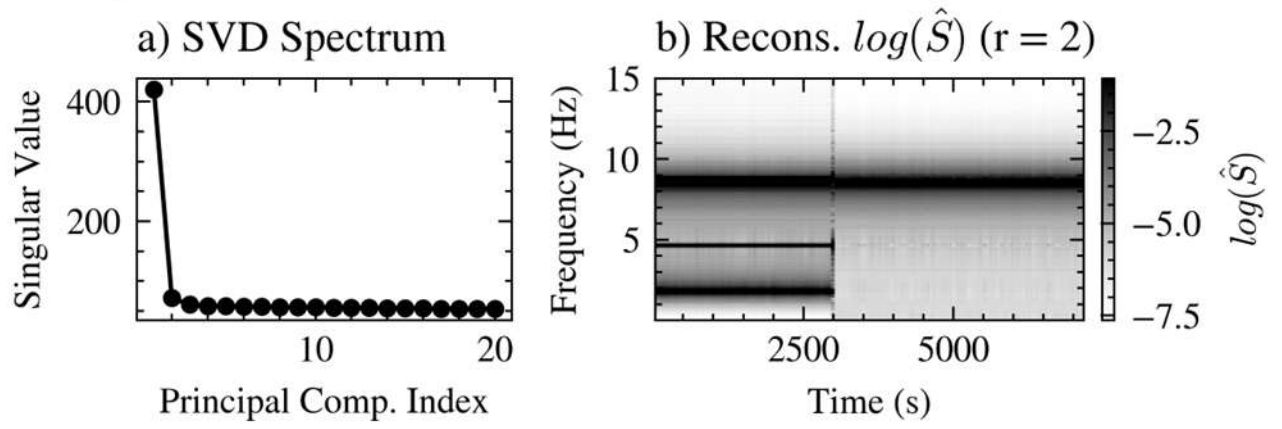

Fig. 3. (a) Singular values of the centered log-periodogram matrix $\log(\hat{S})$ (b) Rank $r$ reconstruction of the log-spectrum using $r = 2$ principal components.

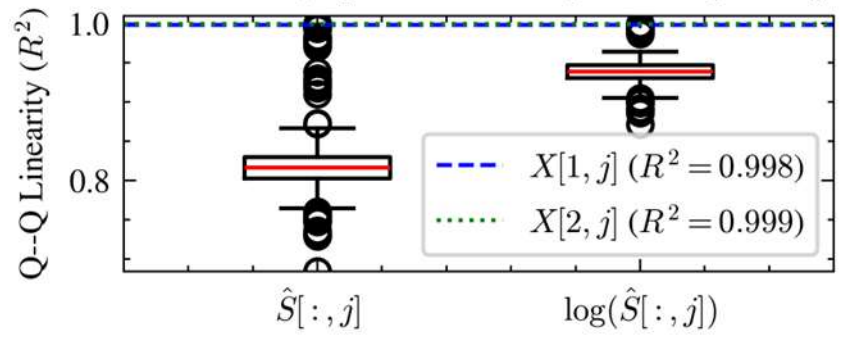

Fig. 4. Distribution of Q-Q plot linearity for raw periodogram $\hat{S}$, its log transform $\log(\hat{S})$ and PCA features $X$.

Fig. 5 plots the cost function from (7) across all 1,000 Monte Carlo realizations, showing minima at $j_1 = 98$ and 100 flanking a local maximum at $j_1 = 99$. This occurs because the spectrogram $\hat{S}[:, j]$ (from (3)) at $j = 98$ and 99 both contain the genuine changepoint at 2984 s, thereby characterized by a mixture of the two stationary regimes at $t < 2984$ and $t \geq 2984$. This results in $j = 98{,}99$ being spectrally different from the stationary regimes $\forall j < 98$ and $\forall j > 99$ (as seen in Fig. 2 and Fig. 3 b)), requiring an additional changepoint at $j = 100$ to model the data. Finally, we evaluate the segmentation performance using the Rand Index [17], which scores sample-pairing accuracy. Across all runs, our detector achieves the best possible solution in $\hat{S}$ space with two breakpoints at ($j_1 = 98, j_2 = 100$), with a high mean score of 0.99, localizing the true transition within 1 block with minor imperfections stemming entirely from the above mentioned boundary effect.

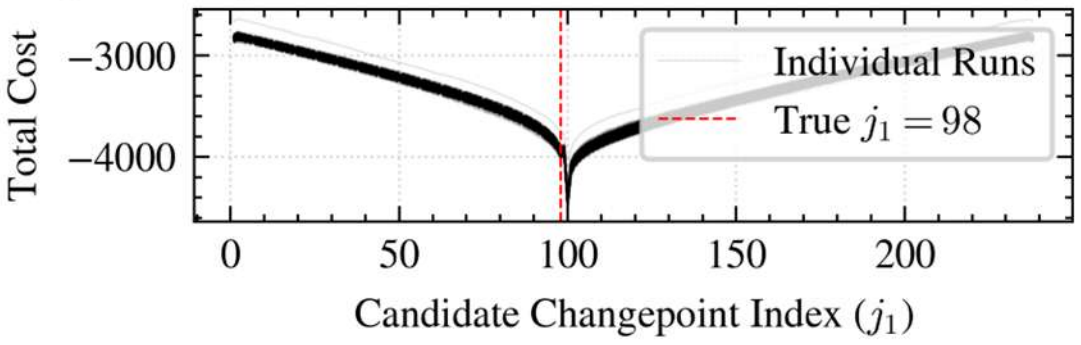

Fig. 5. Cost function evaluated across candidate changepoint indices $j_1$ for 1000 independent runs (black curves), with the true changepoint in red.

### B. PMU Data from Dominion Energy

We now apply the proposed approach to Dominion Energy PMU data from two scenarios to demonstrate its effectiveness in capturing changes in real-world dynamics. Because real-world spectra from large systems can be cluttered, these spectrograms are plotted with a clipped color range to accentuate prominent behaviors and allow visual validation.

The first case is the 100-MW AI training load from Fig. 1, previously analyzed in [6]. While the dominant component of the AI training duty cycle is slow (< 1 Hz), its non-sinusoidal nature introduces a range of harmonics and is also accompanied by other fast dynamics. Therefore, we work with the full spectrum for an interesting 3-hr. window from mid-July 2025 (Fig. 6). Two prominent processes appear: an intermittent one extending up to 7 Hz, and relatively low frequency persistent one. As shown in Fig. 6, the algorithm correctly identifies 10 changepoints that bound all 5 major active bursts of the intermittent 1-7 Hz process, with zero false splits during quiescent intervals. A natural next step is to cluster similar segments, i.e., those sharing the same underlying processes and hence similar spectra; this is left for future work.

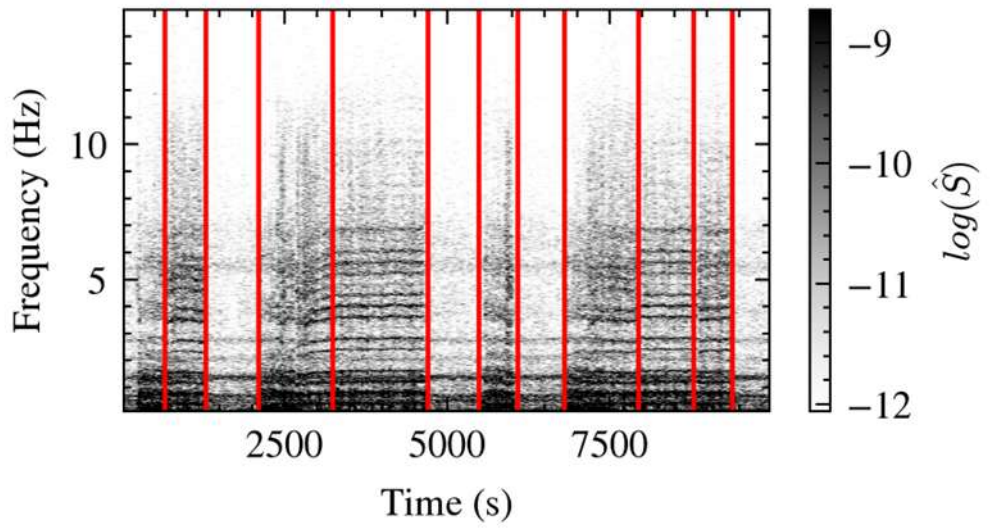

Fig. 6. Real-world AI training load current spectrogram showing a likely combination of two processes, with estimated changepoints (red).

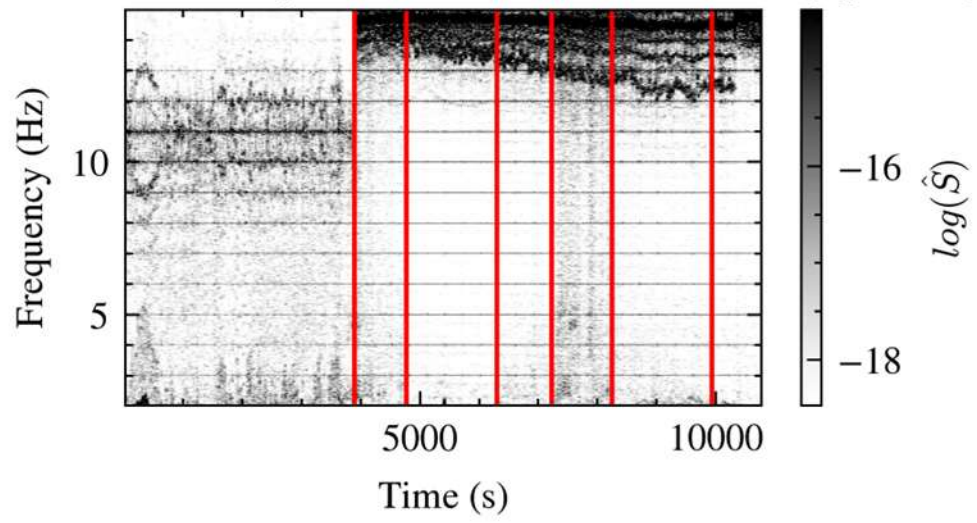

Fig. 7. Data center voltage spectrogram showing > 14 Hz oscillations lasting over 2 hrs., along with other dynamics, with estimated changepoints (red).

The second case is a data center (not an AI training load) in which an unstable UPS caused sustained oscillations lasting over 2 hrs. (analyzed in [1]). The voltage magnitude spectrogram for a 3-hr. data window containing the inception of oscillations is shown in Fig. 7. The horizontal lines at 1-Hz intervals are due to sustained periodic voltage sags, which, being stationary, do not contribute to the dominant principal components ($X$). The first prominent feature is oscillatory behavior centered around 10.5 Hz, which vanishes and gives way to the > 14 Hz oscillations at around 3600 s. This is the first prominent change point, which the proposed approach correctly estimates. The oscillation itself is non-stationary, and its characteristics vary drastically with the grid operating point, as seen in Fig. 7, where its bandwidth progressively widens from 1 Hz to almost 3 Hz. As a result, the approach identifies multiple changepoints along the way, albeit only a few, thereby showing robustness to this departure from the piecewise-stationary spectrum assumed in Section II. Extending the method to more complex signal models is therefore a promising direction. Finally, it also identifies a temporary period of broadband dynamics at around 7000–8000 s, attributed to a nearby electric arc furnace.

## V. Conclusion

In this work, we propose a non-parametric framework to segment ambient synchrophasor data into spectrally homogeneous intervals, enabling the identification of wideband regime shifts in AI training loads and underlying system dynamics in general. By establishing the statistical properties of log-spectral PCA under a mixed-spectrum model, we prove convergence to a diagonal-covariance Gaussian subspace via the Central Limit Theorem. This enables an efficient, BIC-penalized GLR changepoint detector. Monte Carlo evaluations on synthetic data and Dominion Energy PMU measurements from data centers confirm the framework's precision. Future work will explore segment clustering and extensions for slowly varying spectral drifts.